\documentclass[twocolumn]{aastex701}

\def \aap  {A\&A}

\def \apjs  {ApJS}
\def \apjl  {ApJL}

\usepackage{subfigure}
\usepackage{graphicx}  
\usepackage{appendix}
\usepackage{xcolor}

\begin{document}

\title{Detecting Neutrino Emission from Supernova Remnants: A Theoretically Motivated Target Catalog}

\author[0000-0002-9386-630X]{Emily Simon}
\affiliation{Department of Astronomy and Astrophysics, University of Chicago,
Chicago, IL 60637, USA}
\email[show]{ersimon@uchicago.edu}

\author[0000-0002-2160-7288]{Rebecca Diesing}
\affiliation{Department of Physics and Columbia Astrophysics Laboratory, Columbia University, New York, NY 10027, USA}
\affiliation{School of Natural Sciences, Institute for Advanced Study, Princeton, NJ 08540, USA}
\email{rrdiesing@ias.edu}

\author[0000-0003-0939-8775]{Damiano Caprioli}
\affiliation{Department of Astronomy and Astrophysics, University of Chicago,
Chicago, IL 60637, USA}
\affiliation{Enrico Fermi Institute, The University of Chicago,
Chicago, IL 60637, USA}
\email{ersimon@uchicago.edu}

\author[0000-0003-0939-8775]{Stephen Sclafani}
\affiliation{University of Maryland, College Park, MD 20742, USA}
\email{ersimon@uchicago.edu}



\begin{abstract}
Galactic supernova remnants (SNRs) are thought to accelerate cosmic rays (CRs) to several PeV energies, but this has yet to be confirmed as general behavior. 
Although several sources show $\sim 100$ TeV $\gamma-$rays, their hadronic origin is uncertain; a matching neutrino signal would provide definitive evidence.
Using insight from the theory of diffusive shock acceleration, we evaluate the spectra and environments of the sample of Galactic SNRs to identify those most likely to be hadronic, categorizing them into a tiered catalog depending on their likelihood to produce neutrinos detectable in the TeV-PeV range.
We then calculate the estimated stacked sensitivity of IceCube for each tier using IceCube's ten-year public data.
Our results suggest that this strategy of stacking SNRs and carefully excluding leptonic sources by using theoretical arguments may allow for a detection of this source class that would otherwise be impossible.
A follow-up analysis of these catalogs using TeV–PeV sensitive neutrino data from IceCube (or similar telescopes like KM3NeT/ARCA) offers the most decisive, near-future test for the hadronic nature of these SNRs and the maximum energies of their CR spectra.

\end{abstract}

\section{Introduction}

Supernova remnants (SNRs) are widely considered the primary sources of Galactic cosmic rays (CRs), potentially responsible for acceleration up to the ``knee", a steepening in the CR spectrum at PeV ($\sim 10^{15}$ eV) energies \citep{drury83, blasi13}. 
Diffusive Shock Acceleration (DSA) at SNR shocks is a natural mechanism for accelerating particles to relativistic energies \citep{krymskii77, axford+77p, bell78a, blandford+78}, yet some uncertainties remain; 
in particular, there is no conclusive evidence that typical SNRs reliably accelerate protons to PeV energies.
Theoretical works \citep[e.g.,][]{bell+13, blasi+15, cardillo+15, cristofari+21,diesing23} emphasize the challenge of accelerating particles beyond $\sim 100$ TeV at typical isolated SNRs, and observations of high energy Galactic sources \citep[e.g.,][]{cao+24_LHAASOCat, HGPS18} tend to show that sources with energies $\gtrsim 100$ TeV are not associated with known SNRs, but usually with leptonic sources like pulsar wind nebulae (PWNe) \citep{amato14}.

A critical challenge in identifying proton acceleration at SNRs lies in disentangling hadronic $\gamma$-ray emission from competing leptonic processes. 
High-energy $\gamma$-rays can be produced by both neutral-pion decay from proton interactions (hadronic), or by mechanisms such as inverse-Compton scattering or relativistic bremsstrahlung (leptonic), requiring spectral analysis and environmental context to determine the dominant emission process \citep[e.g.,][]{drury+94, morlino+09, ellison+07, corso+23}.
As the CR energy spectrum requires a Galactic population of hadronic accelerators up to PeV energies,
disentangling whether $\gamma$-ray emission is hadronic or leptonic is critical to resolving the long-standing question of the origin of Galactic CRs.

Neutrinos have the unique property of being unambiguous evidence of high-energy proton acceleration, as they must originate from hadronic interactions in conditions typical for SNRs. 
Isolated SNR sources are expected to possess neutrino fluxes too low to be detected by neutrino telescopes like IceCube \citep{IceCube2020}, but stacking analyses--- where the signal from many isolated sources is combined into one signal--- may allow us to detect faint sources in aggregate.
Previous stacked neutrino searches for Galactic sources have found null results \citep{aartsen+17m}, but these samples may also have suffered from contamination of leptonic sources, which can often dominate the $\gamma-$ray contribution, but provide no neutrinos, essentially increasing the combined noise but adding nothing to the combined signal. 
Here we use the current theory of particle acceleration at shocks \citep{haggerty+20, caprioli+20, diesing+21} to build catalogs of only those SNRs which have a strong possibility of being hadronic, thus maximizing the signal to noise in a neutrino stacking analysis.

This work is organized as follows: in Section \S\ref{section:spectral_class}, we discuss the theoretical background for our SNR classification scheme. 
In Section \S\ref{section:data}, we give an overview of the available SNR data that was used in this analysis and describe the procedure for categorizing SNRs and fitting their spectra.
Section \S\ref{section:results} shows the estimated neutrino flux for SNRs in different catalogs compared with the estimated sensitivity of IceCube. 
Finally, in Section \S\ref{section:conclusion} we conclude with prospects for neutrino detection of our catalogs and implications for SNRs as the main source of Galactic CRs. 

\section{Spectral Classification}
\label{section:spectral_class}

The photon spectrum of an SNR contains information about the physical processes producing the radiation.
The ``smoking gun" of hadronic interactions is the pion decay bump, a spectral feature which results from neutral pions decaying around their rest mass energy $\sim 135$ MeV \citep[e.g.,][]{drury+94}. 
The decay produces two $\gamma$-rays, each of which carry roughly half of the pion's energy, leading to a $\gamma$-ray bump around $67.5$ MeV.
This bump, which for a power-law distribution of hadrons shows up as a cut-off below $\sim 100$ MeV, does not have a leptonic analog and is therefore a conclusive hadronic signature;
it has been observed by Fermi-LAT in sources such as W44 and IC443 \citep{ackermann+13, cardillo+14, liu+24}, but in most SNRs the sub-GeV flux is too low to reveal this characteristic feature.

However, GeV-TeV observations also encode information about the origin of an SNR's $\gamma$-rays; in particular, Fermi-LAT provides spectral indices of SNRs across this energy range, which can be used to infer the emission mechanism 
\citep{caprioli11,caprioli12}.
In general, diffusive shock acceleration (DSA) produces energy power-laws $dN_{\gamma}/dE_{\gamma}\propto E_\gamma^{-q}$, with $q = (r+2)/(r-1)$, where $r$ is the compression ratio of the shock and depends only on the shock Mach number \citep{bell78a}.
For strong SNR shocks, $r \to 4$, and thus $q=2$, though corrections due to self-generated magnetic fields suggest that values $q\sim 2.2-2.4$ are actually realized at strong shocks \citep{haggerty+20, caprioli+20}. 
Indeed, observations of individual SNRs which are widely believed to be hadronic (e.g., W44, W28, IC 443) show spectra steeper than $E^{-2}$ \citep[e.g.,][]{w44Fermi, W28Fermi, IC443Fermi}.

Inverse Compton, the upscattering of photons off of relativistic electrons, is the dominant leptonic radiation process and produces a hard $\gamma$-ray spectral index $\propto -(q+1)/2$, where $q$ is the electron slope. 
Acceleration via DSA is rigidity-dependent, meaning the electron slope should be the same as the proton slope at low (sub $\sim$ TeV) energies where synchrotron losses are negligible \citep{ellison+91, amato+06}.
For the reference $q \sim 2$ electron spectrum, the inverse-Compton $\gamma$-ray energy spectrum scales as $dN_\gamma/dE_\gamma\propto E_\gamma^{-1.5}$. 
Conversely, $\pi_0$ decay produces photon spectra parallel to the parent spectrum, i.e., a $E^{-2}$ proton spectrum would yield a photon spectrum $dN_\gamma/dE_\gamma\propto E_\gamma^{-2}$, appreciably softer than the leptonic one.
Finally, relativistic bremsstrahlung \citep{blumenthal+70} produces a spectrum parallel to that of the parent electrons, but it is almost invariably under-dominant with respect to either inverse-Compton or pion production in low or high density interstellar mediums, respectively.

This information allows us to classify SNRs by their spectral hardness, identifying candidates that are highly likely to be dominated by hadronic emission.
Broadly speaking, DSA spectra have $2\lesssim q\lesssim 3$, so any photon spectrum softer than $E_\gamma^{-2}$ must be hadronic (an inverse-Compton spectrum would require an unphysical $q\gtrsim 3$), and any spectrum harder than $E_\gamma^{-2}$ is most likely to be leptonic.
Importantly, spectra are evaluated over the GeV-TeV range, but not beyond, to avoid probing the spectral cutoff which can mimic the appearance of steep spectra typical of hadronic sources, but for unrelated reasons.

Mechanisms for spectral hardening of hadronic SNRs from interactions with clumpy interstellar media have also been proposed \citep[e.g.,][]{celli+19, gabici+14, zirakashvili+10}, where amplified magnetic fields around the clumps prevent low-energy CRs from entering and contributing to the $\gamma$-ray flux, but further work is necessary to quantify the ubiquity of such a mechanism and the fine-tuning of the ensuing spectra.
Nevertheless, we note that erroneously \textit{excluding} a hadronic source will typically be less punishing in a stacked analysis than erroneously \textit{including} a leptonic source. 
Our analysis can be interpreted as taking a ``conservative" approach where it is possible that some hadronic SNRs are missing, but very unlikely that any leptonic sources are included.

\section{Selections of Data}
\label{section:data}

\begin{figure}[h!]
    \centering
    \includegraphics[width=0.99\linewidth]{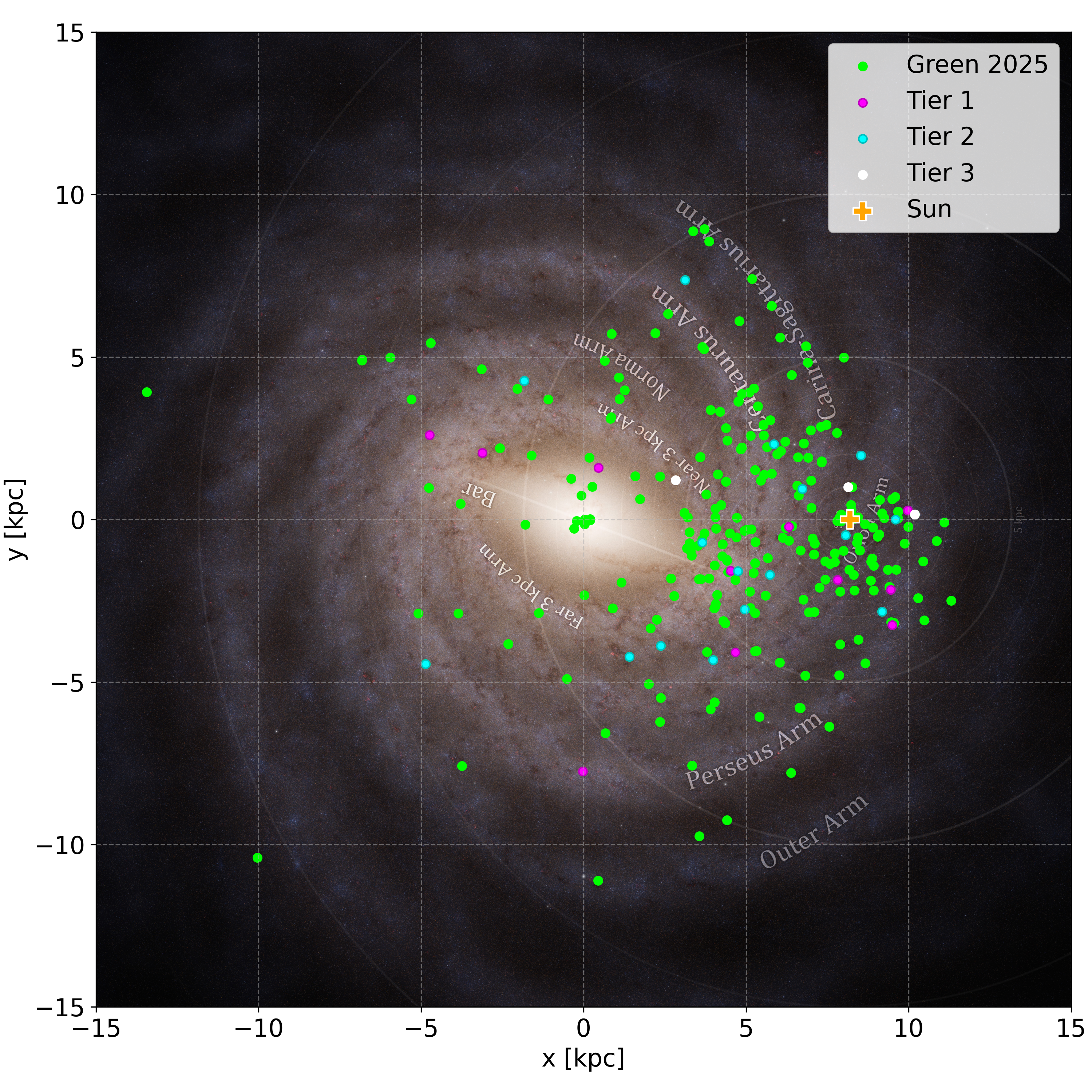}
    \caption{The relative positions of known Galactic SNRs from \cite{Green25}(green points), and the selected subset of SNRs in our Tier 1, 2, and 3 catalogs (magenta, cyan, and white points respectively). Source for background image: Gaia [ESA/Gaia/DPAC, Stefan Payne-Wardenaar] \cite{GaiaMilkyWayMap2025}. Distances are approximate and are taken from \cite{Green25} when available, and otherwise from \cite{Ranasinghe+24} and \cite{Wang+20} to reach complete coverage for all SNRs.}
    \label{fig:MW_SNRs}
\end{figure}

From the sample of $310$ known Galactic SNRs \citep{Gilles+12, Green25}, we have limited our focus to those with high-energy spectral observations.
Of this total population, Chandra's SNR catalog (\url{https://hea-www.harvard.edu/ChandraSNR/snrcat_gal.html}) contains $152$ observed in X-rays, and Fermi's First Galactic SNR catalog \citep{FermiSNRCat16} further reduces this number to 30 known SNRs with several additional candidates.
The TeVCat catalog \citep{TeVCat} lists 56 TeV-bright Galactic sources non-exclusively typed as SNRs, and overlaps with a portion of those in the Fermi-LAT catalog.
Additionally, our samples have also been supplemented with various observations from HAWC, HESS, VERITAS, MAGIC, Tibet AS$\gamma$, and LHAASO where available (see tables for full list of references). 

\subsection{Tiers 1, 2, and 3}
We categorize the SNRs of interest into three tiers based on their likelihood of being hadronic and their expected detectability by IceCube, which is only sensitive above the atmospheric background at energies around a TeV or greater \citep{IceCube2020}. 

\begin{itemize}
    \item {\bf Tier 1}: SNRs that are extremely likely to be hadronic and potentially observable with IceCube above the atmospheric neutrino background.
    \item {\bf Tier 2}: SNRs that are likely to be hadronic, but it is inconclusive whether their neutrino emission will extend to energies above IceCube's lower energy threshold. 
    \item {\bf Tier 3}: SNRs that are leptonically dominated, but bright enough to potentially contain an underdominant hadronic component with flux comparable to sources in Tier 1.
\end{itemize}

\begin{figure*}[t]
    \centering
    \includegraphics[trim = 100 25 100 30, clip=true, width=0.98\linewidth]{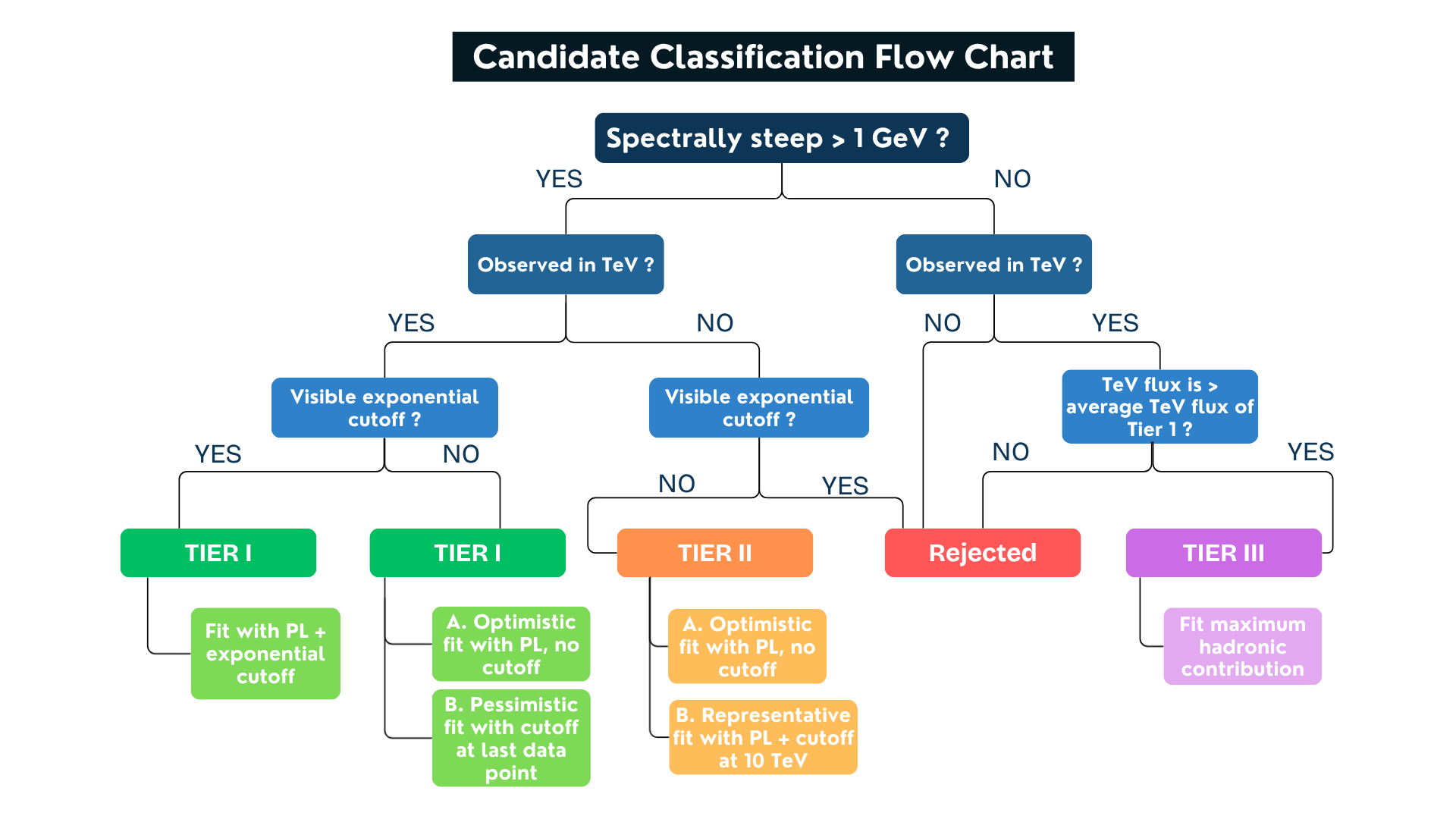}
    \caption{Flowchart of our classification procedure to sort SNRs into Tiers 1, 2, or 3, or to exclude them from all tiers. Steep spectra are defined as those with spectral indices $q>2.1$. Beneath each tier is the tier-dependent analysis strategy for evaluating the neutrino contribution of the contained sample of SNRs, where PL stands for power law, either with or without an exponential cutoff as noted.}
    \label{fig:flowchart}
\end{figure*}

Figure \ref{fig:MW_SNRs} shows the approximate positions of all known Galactic SNRs super-imposed on a reconstruction of the Milky Way as seen by Gaia \citep{GaiaMilkyWayMap2025}.
The approximate positions of our selected sources for Tiers 1, 2, and 3 are in magenta, cyan, and white points respectively.
For visual consistency, distances have all been estimated following the convention: adopt the value from \cite{Green25} if it exists; if not, adopt the value from \cite{Ranasinghe+24} and then from \cite{Wang+20}. 

We build our Tier 1, 2, and 3 catalogs by examining the available literature for all SNRs known to have $\gamma-$ray emission.
From the literature, we do a first evaluation of the spectral data in the GeV-TeV band, looking for spectral indices which have $q > 2.1$, our threshold for considering spectra to be steep and likely associated with ion acceleration via DSA.
For these sources, we compile all available spectral data over all available instruments from GeV energies and beyond, prioritizing data from more recent studies if multiple datasets exist for the same instrument. 
We then check whether each source's spectrum extends out to TeV energies; those which unambiguously extend to TeV energies are added to Tier 1.
Those which do not have observed data beyond a TeV, but which have no sign of a spectral cutoff in the observed GeV data, are added to Tier 2 under the optimistic assumption that their spectra continue to TeV energies, though we lack the observational coverage to confirm it.
SNRs which are spectrally steep, do not have observed TeV emission, but do have an observed spectral cutoff at energies below a TeV are excluded from all tiers as they will fall below IceCube's sensitivity threshold. 
SNRs found in the literature which are $\gamma-$ray bright, extend to TeV energies, but have spectra harder than $q = 2.1$ are likely to be leptonic in nature and are thus tentatively added to Tier 3, but can only be confirmed in this tier after their possible neutrino spectrum is modeled. 
Those with a modeled neutrino spectrum that is comparable to the average expected neutrino flux of sources in Tier 1 are kept in Tier 3, while those with significantly smaller fluxes are excluded from all tiers.
A flowchart of our categorization procedure can be found in Figure \ref{fig:flowchart}, and the lists of Tier 1, 2, 3, and excluded SNRs (and their reason for exclusion) can be found in Tables \ref{tab:Tier1}-\ref{tab:exclusions} respectively.
A visual comparison of the spectra for a representative SNR from each Tier of the catalog is shown in Figure \ref{fig:spec_comps}, where W51C (purple) is a typical Tier 1 source, W44 and MGRO J1908+06 (blue and orange) are both typical Tier 2 sources, and Vela Jr (green) is typical Tier 3 source.

\subsection{Estimated neutrino flux per source}

After categorizing relevant SNRs into their respective tiers (or tentatively categorizing them for Tier 3), we performed a basic spectral fit on the data for each individual source. 
These fits were intentionally kept simple, utilizing only the central flux values and omitting error bars and upper limits, since the resulting spectral parameters are intended only as input values for our subsequent stacked analysis. 
The parameters from each fit are used to estimate the flux in neutrinos from each source and assuming that that $dN_\gamma/dE \simeq dN_\nu/dE$ for proton-proton collisions \citep{kelner+06}, and the procedure for this estimate varies depending on the tier.

\begin{figure}
    \centering
    \includegraphics[trim = 0 0 0 0, clip=true, width=0.98\linewidth]{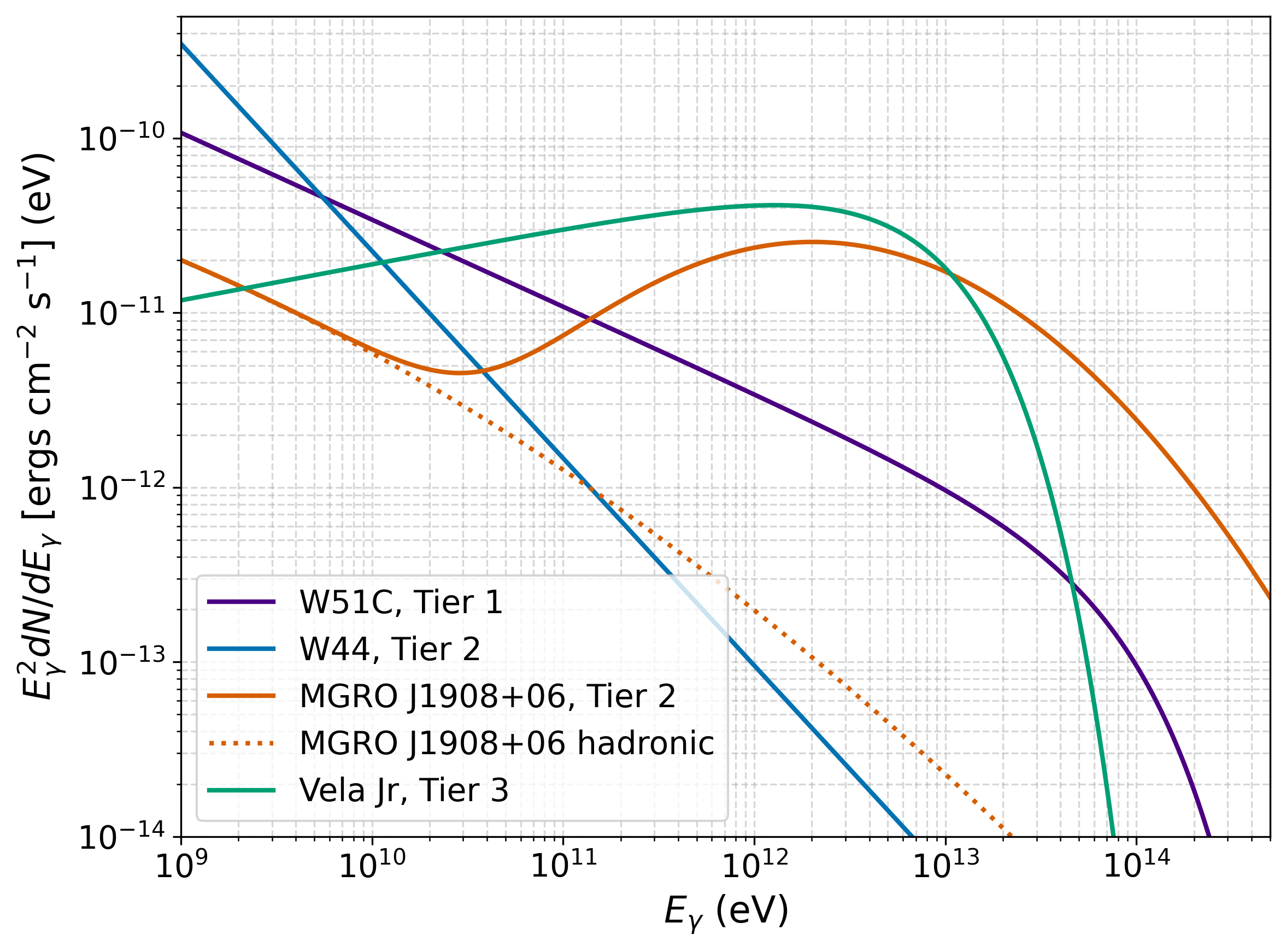}
    \caption{Spectral fits of representative SNRs from each Tier of our catalog: W51C (Tier 1, purple) has a steep spectrum and observed emission at energies above a TeV; W44 (Tier 2, blue) has a steep slope but has not been observed above a TeV; MGRO J1908+06 (Tier 2, orange) has a concave spectrum where the low-energy portion may be hadronic and may extend subdominantly to TeV energies (orange dotted line); Vela Jr (Tier 3, green) has a hard spectral index but is very bright at TeV energies.}
    \label{fig:spec_comps}
\end{figure}

\begin{figure}
    \centering
    \includegraphics[width=0.98\linewidth]{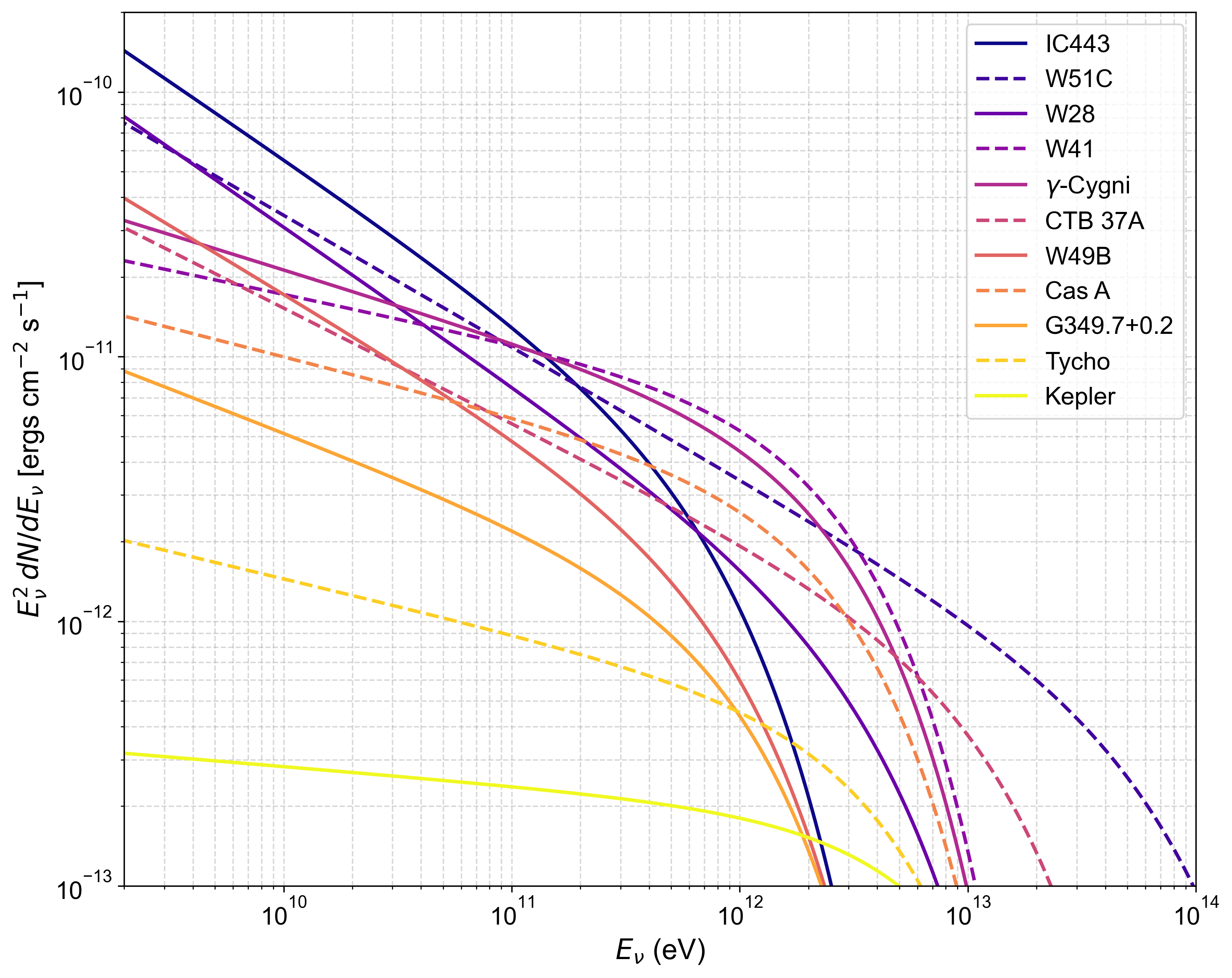}
    \caption{Spectral fits for all sources in the Tier 1 catalog assuming power laws with exponential cutoffs.}
    \label{fig:Tier1_exp}
\end{figure}

Tier 1 contains a mixture of SNRs with and without visible exponential cutoffs.
Those with visible cutoffs are fit assuming a power law from energies $\geq$ GeV and an exponential cutoff,
\begin{equation}
    \frac{dN}{dE} = AE^{-\alpha}\exp\left[{E/E_{\rm cut}}\right]
    \label{PL+exp}
\end{equation}
where A is the normalization, $\alpha$ is the spectral index at energies $\geq$ GeV, and $E_{\rm cut}$ is the cutoff energy.
Those without visible cutoffs are fit in two scenarios, one ``optimistic" and one ``pessimistic".
The optimistic case fits the power law $\geq$ GeV with $E_{\rm cut}\to \infty$, representing the (naive) best case scenario for the neutrino flux. 
The pessimistic case fits the power law and exponential cutoff, as in Eq. \ref{PL+exp}, but sets $E_{\rm cut}$ manually to be the energy of the last observed data point (excluding upper limits). 
These fits together bracket the range of possible neutrino fluxes expected from a given source.
The expected neutrino flux for all Tier 1 sources assuming a power law with an exponential cutoff is shown in Figure \ref{fig:Tier1_exp}. 
The fits with $E_{\rm cut} \to \infty$ for the $5$ sources without observed spectral cutoffs are not shown for sake of clarity, but the best fit parameters are included in Table \ref{tab:Tier1} in rows labeled (b).

The sources (or hadronic components of sources) in Tier 2 all have observed spectra below a TeV, and fitting them pessimistically with exponential cutoffs at their last observed data point would preclude them from emitting neutrinos within IceCube's sensitivity range.
The most optimistic case with no cutoff is interesting, as it provides an upper limit to the expected neutrino flux from these sources which can then be compared to IceCube's sensitivity. 
However, SNRs are virtually guaranteed to have a spectral cutoff, so we also consider the Tier 2 sample with a representative cutoff energy approximately equal to the average cutoff energy for our Tier 1 sample, $10$ TeV.
Therefore we obtain a power law fit for all Tier 2 sources assuming $E_{\rm cut} \to \infty$, but use both prescriptions for the cutoff energy (or lack thereof) when calculating the IceCube sensitivity in subsequent steps.

\begin{figure}
    \centering
    \includegraphics[width=0.98\linewidth]{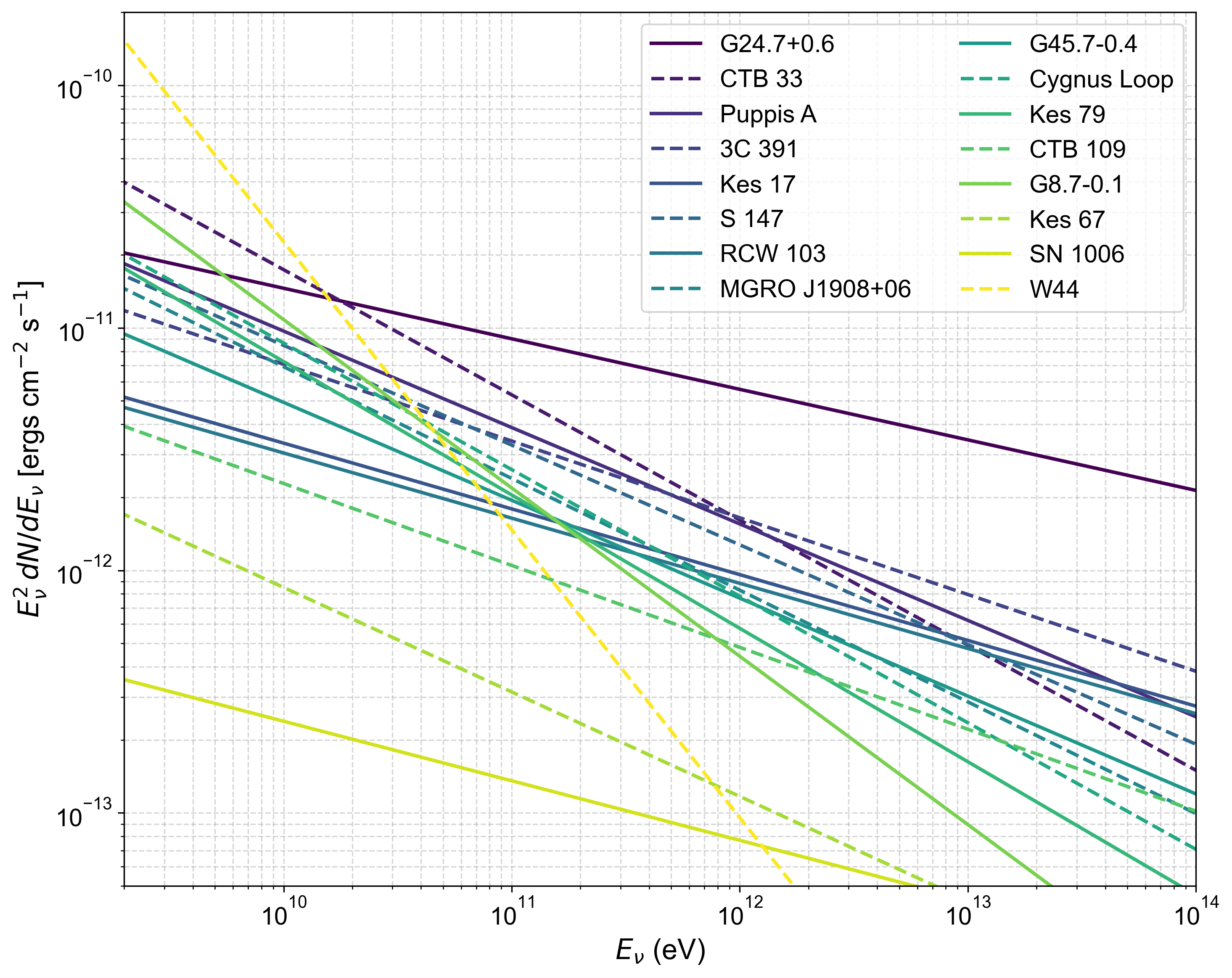}
    \caption{Spectral fits for all Tier 2 sources assuming power-law spectra with no cutoffs.}
    \label{fig:Tier2_PL}
\end{figure}

Three sources in Tier 2 have concave (or potentially concave) spectra in the GeV-TeV range: MGRO J1908+06 \citep{MGRO1908_Fermi21, MGRO1908_Fermi+Veritas+HAWC24}, CTB 109 \citep{CTB109_Fermi12, CTB109_Fermi23}, and potentially the southwest limb of SN 1006 \citep{lemoine-goumard+25}.
For these sources, it has been suggested in the aforementioned studies that the spectra are best modeled with a combined lepto-hadronic model. 
Specifically, the spectral peak around $\sim1$ GeV is due to $\pi^0$ decay which is overtaken by a rising leptonic contribution at higher energies, ultimately leading to a concave spectrum (see Figure \ref{fig:spec_comps}, orange solid line).
We perform a fit to only the assumed hadronic portions of these three sources, assigning a power law to the emission $\geq$ GeV but \textit{excluding} data points where the spectrum begins to rise again around tens of GeV (see Figure \ref{fig:spec_comps}, orange dotted line). 
This allows us to extrapolate the assumed hadronic portion of the spectrum to energies $\gtrsim$ a TeV for comparison to IceCube sensitivities.
The power law fits for all Tier 2 sources are shown in Figure \ref{fig:Tier2_PL}, and all fit parameters can be found in Table \ref{tab:Tier2}.

Tier 3 is comprised of sources with spectra harder than $E^{-2}$, where the bulk of $\gamma-$ray emission is most likely coming from PWNe and thus is very likely to be leptonically-dominated. 
However, these sources may also contain an under-dominant hadronic contribution. 
For particularly bright sources, relatively large hadronic contributions could be hidden beneath leptonic spectra, but still contribute neutrinos with comparable fluxes to sources in Tier 1.
For Tier 3, we fit the total spectrum using either a log-normal (e.g., the Crab) or power-law with exponential cutoff (e.g., Vela Jr), where appropriate to capture the shape of the spectrum, remaining agnostic to the precise physics behind the shape.
We then add a ``representative" hadronic contribution with spectral index $dN/dE \propto E^{-2}$ assuming an ideal DSA-accelerated population of protons, such that the sum of the hadronic contribution and the leptonic fit do not exceed $10\%$ of the original leptonic fit, ie:
\begin{equation}
    S_{\rm fit}(E) + S_{\rm hadronic}(E) \leq 1.1 S_{\rm fit}(E).
\end{equation}
We then solve for values of the normalization, A, and $E_{\rm cut}$ which maximize the integrated spectrum of the hadronic component $\geq 1$ TeV.
This gives us an estimate of the maximum hadronic contribution that could be effectively hidden beneath the leptonic spectrum. 
We then compare whether this optimal integrated flux is greater or equal to the average source in Tier 1, which is found to be $\sim 1.33\times10^{-12}$ TeV/cm$^2$/s. 
Sources with optimal hadronic fluxes that are smaller than the average flux of a Tier 1 source are excluded.
Those with optimal hadronic fluxes comparable to or larger than the average in Tier 1 are added to Tier 3. 
A table of which sources are included in Tier 3 and their maximal hadronic contribution (assuming an $E^{-2}$ spectrum) can be found in Table \ref{tab:Tier3}. 
The full list of excluded SNRs and the reason for exclusion can be found in Table \ref{tab:exclusions}.

\subsection{Notes on Individual Sources}

\subsubsection{Cas A}
A recent study of Cassiopeia A (Cas A) from \cite{CasA_LHAASO25} has shown a slightly concave spectrum in the GeV-TeV range in data from Fermi-LAT, similar to some of the sources in our Tier 2 catalog.
The authors suggest there is a possibility that this may be due to a low energy hadronic component which is overtaken by a leptonic component at higher energies. 
Beyond several tens of TeV, the leptonic component then drops off steeply and the hadronic component takes over once more. 
In this modeling, the true hadronic component would have a significantly steeper spectrum than what was found in this work where we assumed the entire spectrum to be hadronic, which will then impact the expected neutrino flux from this source.

\subsubsection{SNR G24.7+0.6}
SNR G24.7+0.6 is in a very complex environment, which makes it difficult to understand the detailed origins of the highest-energy emission. 
Previous work by \cite{G24.7_Fermi+MAGIC19} combined Fermi-LAT and MAGIC observations and found good spectral agreement and overlapping source positions within 1.5$\sigma$.
Including the MAGIC contribution would extend the spectrum out to $>$ TeV energies and show evidence of an exponential cutoff, making G24.7+0.6 a good candidate for our Tier 1 catalog.
However, if the MAGIC source were not the same as the Fermi-LAT one, G24.7+0.6 would have no confirmed emission above a TeV and be therefore better suited for Tier 2, meaning it is likely hadronic but there is some uncertainty to its detectability by IceCube.
To err on the side of caution, we have categorized the source as Tier 2 and excluded the MAGIC observations from our fit, but future observations of G24.7+0.6 may provide reasons to upgrade it to Tier 1.

\subsubsection{G045.7-00.4}
The energy spectrum of SNR G045.7-0.4 presented in \cite{G45.7_Fermi21} using Fermi-LAT data shows data points out to tens of GeV, and upper limits out to hundreds of GeV. 
Beyond this, the spectrum is extrapolated and compared to the predicted sensitivity curve of LHAASO assuming one year of point source exposure \citep{LHAASO_scienebook19}, which demonstrates that G045.7-0.4 should be detectable by LHAASO if the source has a continuous power-law spectrum at energies above a TeV.
However, LHAASO's first catalog \citep{cao+24_LHAASOCat} does not include a firm association with this source, although it does include an association for 1LHAASO J1914+1150u with 2HWC J1914+117 (with a 0.13 degree spatial separation).
This source lies within 1 degree of G045.7-0.4 (roughly 0.67 degrees apart using the SNR location from \cite{Green25}). 
The current evidence is ambiguous as to whether this is extended emission related to G045.7-0.4, or whether there is no relation, G045.7-0.4 is missed by LHAASO, and thus the SNR must have a spectral cutoff likely around a few TeV or less.
Due to the current ambiguity, we have chosen to keep the source in our Tier 2 catalog with an optimistic fit, but further observations may provide reason to downgrade this source into the excluded catalog due to its incompatiblity with IceCube detection. 

\subsubsection{Vela Jr and RX J1713.7}
Vela Jr (RX J0852.0-4622) is the brightest source in our sample, and has an integrated flux over $1$ TeV slightly larger than the flux of the Crab.
Its spectral slope is quite hard, with $\alpha = 1.81 \pm 0.08$ \citep{HGPS18}.
A slope this flat due to hadronic interactions cannot be well-interpreted in the framework of standard DSA, which invariably produces spectra steeper than $E^{-2}$, and this suggests that Vela Jr is most likely dominated by leptonic emission. 
However, we note that Vela Jr has several distinct $\gamma$-ray bright regions \citep{VelaJr_HESS18} and possible associations with shocked molecular clouds \citep{pakhomov+2012, fukui+17}, which can provide the necessary targets for proton-proton collisions and typically lead to hadronic spectra \citep{corso+23}.
It has been suggested that the spectrum of CRs interacting with molecular clouds may become harder than $E^{-2}$ for a variety of reasons \citep{gabici+09, celli+19, inoue19}, which may explain the contradictory indicators in Vela Jr, and imply that it is indeed a hadronic source.
In order to err on the side of caution, we assume that Vela Jr is predominately leptonic, but include it in our Tier 3 catalog as a source which may contribute neutrinos from an under-dominant hadronic component.

Similarly, RX J1713.7 is one of the brightest known Galactic SNRs in $\gamma-$rays, and has a rather hard spectral index of $\alpha  = 2.06 \pm 0.02$ \citep{HGPS18}, indicating that it is a leptonic source.
Like Vela Jr, however, RX J1713.7 may be associated with several molecular clouds \citep{sano+13}, and has been suggested to have mixed lepto-hadronic emission \citep{fukui+21}.
As with Vela Jr, we have assumed the majority of emission from RX J1713.7 is leptonic, but that an under-dominant hadronic component may contribute toward neutrino production.
Both SNRs are treated here with a conservative approach within the perspective of traditional DSA, but it is possible that further theoretical or observational developments will conclusively favor moving them to a different tier catalog.

\subsubsection{CTA 1}
Recent work from \cite{CTA1_LHAASO25} shows the high energy spectrum of CTA 1 from combined measurements using Fermi-LAT, VERITAS, and both of LHAASO's WCDA and KM2A detectors. 
The spectrum appears to have a concave feature across the VERITAS band, which is somewhat reminiscent of other SNRs like MGRO J1908+06 \citep{MGRO1908_Fermi+Veritas+HAWC24} or the recent observations of Cas A as mentioned above \citep{CasA_LHAASO25}. 
For these SNRs, it can be argued that this concave feature is due to a lower-energy hadronic component which is dominant around a GeV, but is overtaken by a leptonic component at higher energies, giving the appearance of two bumps in the spectrum.
This argument is supported by relatively steep spectral indices from $\sim 1-10$ GeV.
However, the lower-energy bump for CTA 1 has a hard spectrum throughout the GeV-TeV range, which indicates that this emission, as well as the higher energy bump above $\sim10$ TeV, are both likely to be leptonic.
This could mean that the spectrum is actually comprised of a convolution of more than one leptonic source, or possibly that the normalization of the VERITAS data is slightly different from that of other telescope collaborations causing the appearance of a dip from $\sim1-10$ TeV. 
In either case, we find that this source is most likely to be leptonic, and thus consider it for Tier 3 but ultimately exclude it from our catalogs.

\begin{table*}[t]
    \centering
    \small
    \begin{tabular}{ l l l l p{3cm} l}
        \multicolumn{6}{c}{\textbf{Tier 1 SNR Fit Parameters}} \\
        \hline \hline
        \textbf{Name} & \textbf{A [eV]} & \textbf{$\alpha_{\rm GeV}$} & \textbf{E$_{\rm cut}$ [eV]} & \textbf{Integrated flux $\geq 1$ TeV [$\rm cm^{-2} s^{-1}$]} & \textbf{References} \\
        \hline
        Cas A & 1.54e-9 & 2.219 & 2.86e12 & 1.61e-13 & 1,2,3 \\
        W51C & 3.28e-6 & 2.498 & 7.71e13 & 3.52e-12 & 4,5 \\
        Tycho & 1.66e-10 & 2.206 & 4.60e12 & 3.57e-13 & 6 \\
        $\gamma-$Cygni & 9.71e-9 & 2.266 & 2.79e12 & 2.59e-12 &7,8 \\
        G349.7+02 & 1.08e-8 & 2.332 & 1.06e12 & 1.01e-13 &9,10,11 \\
        IC443 & 4.03e-5 & 2.586 & 8.17e11 & 1.61e-13 &12,13,14 \\
        Kepler & (a) 1.47e-12  & (a) 2.071  & 8.38e12 & (a) 1.89e13 & 15,16,17\\
        &(b) 1.10e-11 & (b) 2.157&&(b) 1.96e-13 & \\
        W28 & (a) 2.89e-5 & (a) 2.597  & 4.12e12 & (a) 9.66e-13 & 18,19 \\
        &(b) 1.59e-4 & (b) 2.672 &&(b) 1.32e-12&\\
        CTB 37A & (a) 3.36e-7  & (a) 2.434 & 1.39e13 & (a) 1.80e-12 &20,21,22 \\
        &(b) 1.45e-6 &(b) 2.499&& (b) 1.59e-12&\\
        W49B & (a) 2.50e-6 & (a) 2.516 & 1.00e12 & (a) 1.20e-13 & 23,24 \\
        &(b) 1.17e-4& (b) 2.685 &&(b) 6.62e-13&\\
        W41 & (a) 1.10e-9 & (a) 2.180 & 2.77e12 & (a) 3.22e-12 & 16,25,26,27 \\
        &(b) 4.55e-8 & (b) 2.338 &&(b) 4.74e-12 &\\
        \hline \hline
    \end{tabular}
    \caption{The best fit normalization (A), spectral index from GeV-TeV ($\alpha_{\rm GeV}$), exponential cutoff ($E_{\rm cut}$), and references used for each SNR in our Tier 1 catalog. Sources W28, CTB 37A, W 49B, W 41, and Kepler are all fit twice, where fit (a) represents the best fit power law with exponential cutoff, and fit (b) represents a single power law with no cutoff. \\
    References: (1) \cite{CasA_Fermi10}; (2) \cite{CasA_MAGIC17}; (3) \cite{CasA_VERITAS20}; (4) \cite{W51C_Fermi16}; (5) \cite{W51C_LHAASO24}; (6) \cite{Tycho_all17}; (7) \cite{GammaCyg_AGILE19}; (8) \cite{GammaCyg_Fermi+VERITAS18}; (9) \cite{G349_Fermi10}; (10) \cite{G349_Fermi2012}; (11) \cite{G349_HESS15}; (12) \cite{IC443_Fermi10}; (13) \cite{IC443_Fermi13}; (14) \cite{IC443_MAGIC07}; (15) \cite{Kepler_Fermi21}; (16) \cite{Kepler_Fermi22}; (17) \cite{Kepler_HESS22}; (18) \cite{W28_Fermi10}; (19) \cite{W28_HESS08}; (20) \cite{CTB37A_Fermi13}; (21) \cite{CTB37A_Fermi20}; (22) \cite{CTB37A_HESS08}; (23) \cite{W49B_Fermi10}; (24) \cite{W49B_HESS+Fermi18}; (25) \cite{W41_Fermi13}; (26) \cite{W41_HESS11_ICRC}; (27) \cite{W41_MAGIC06}.}
    \label{tab:Tier1}
\end{table*}

\begin{table*}[t]
    \centering
    \small
    \begin{tabular}{l l l p{3cm} l }
        \multicolumn{5}{c}{\textbf{Tier 2 SNR Fit Parameters}} \\
        \hline \hline
        \textbf{Name} & \textbf{A [eV]} & \textbf{$\alpha_{\rm GeV}$} & \textbf{Integrated flux $\geq 1$ TeV [$\rm cm^{-2} s^{-1}$]} & \textbf{References} \\
        \hline
        G24.7+0.6 & 1.77e-9 & 2.208 & 7.40e-12 & 1,2,3 \\
        CTB 33 & 2.52e-6 & 2.516 & 1.71e-12 & 2,4 \\
        Puppis A & 9.18e-8 & 2.398 & 1.78e-12 & 5,6 \\
        3C 391 & 1.04e-8 & 2.317 & 2.01e-12 & 7,8 \\
        Kes 17 & 1.72e-9 & 2.271 & 1.21e-12 &9 \\
        S 147 & 1.10e-7 & 2.411 & 1.45e-12 & 8,10 \\
        RCW 103 & 1.48e-9 & 2.269 & 1.12e-12 & 11\\
        MGRO J1908+06 & 2.80e-7 & 2.461 & 9.09e-13 & 12,13 \\
        G45.7-0.4 & 5.37e-8 & 2.404 & 8.76e-13 & 14\\
        Cygnus Loop & 1.43e-6 & 2.522 & 8.24e-13 & 8\\
        Kes 79 & 2.30e-6 & 2.550 & 5.96e-13 & 8,15 \\
        CTB 109 & 5.44e-9 & 2.338 & 5.77e-13 & 8,16,17\\
        G8.7-0.1 (W30) & 9.44e-9 & 2.694 & 4.18e-13 & 18 \\
        Kes 67 & 1.73e-8 & 2.431 & 1.31e-13 & 19\\
        W44 & 1.70e1 & 3.188 & 6.99e-14 & 20 \\
        SN 1006 & 6.82e-11 & 2.246 & 6.62e-13 & 21,22,23 \\
        \hline \hline
    \end{tabular}
    \caption{The best fit normalization (A) and spectral index ($\alpha_{\rm GeV}$) for $E \geq$ GeV for each SNR in our Tier 2 catalog. \\
    References: (1) \cite{acero+16}; (2) \cite{G24.7_Fermi17} ; (3) \cite{G24.7_Fermi+MAGIC19}; (4) \cite{CTB33_Fermi13}; (5) \cite{PuppisA_Fermi25};(6) \cite{PuppisA_HESS15}; (7) \cite{3C391_Fermi14}; (8) \cite{suzuki+22}; (9) \cite{Kes17_Fermi23}; (10) \cite{S147_Fermi24}; (11) \cite{RCW103_Fermi24}; (12) \cite{MGRO1908_Fermi21}; (13) \cite{MGRO1908_Fermi+Veritas+HAWC24}; (14) \cite{G45.7_Fermi21}; (15) \cite{Kes79_Fermi14}; (16) \cite{CTB109_Fermi12}; (17) \cite{CTB109_Fermi23}; (18) \cite{W30_Fermi19}; (19) \cite{Kes67_Fermi25}; (20) \cite{ackermann+13}; (21) \cite{SN1006_fermi25}; (22) \cite{SN1006_fermi19}; (23) \cite{SN1006_HESS10}.}
    \label{tab:Tier2}
\end{table*}

\begin{table*}[t]
    \centering
    \small
    \begin{tabular}{l l l p{3cm} l }
        \multicolumn{5}{c}{\textbf{Tier 3 Hadronic Component Fit Parameters}} \\
        \hline \hline
        \textbf{Name} & \textbf{A [eV]} & \textbf{$E_{\rm cut}$ [eV]} & \textbf{Integrated flux $\geq 1$ TeV [$\rm cm^{-2} s^{-1}$]} & \textbf{References} \\
        \hline
        Crab & 3.30e-12 & 6.70e12 & 5.29e-12 & 1,2,3,4,5\\
        Vela Jr & 1.18e-12 & 1.03e13 & 1.89e-12 & 6,7 \\
        RX J1713.7 & 4.39e-13 & 4.14e13 & 7.04e-13 & 8,9,10\\
        \hline \hline
    \end{tabular}
    \caption{The best fit normalization (A) and $E_{\rm cut}$ for the maximal hadronic contribution of sources in our Tier 3 catalog assuming an $E^{-2}$ spectrum.\\
    References: (1) \cite{Crab_Fermi+HESS24} (2) \cite{Crab_Fermi12}; (3) \cite{Crab_MAGIC15}; (4) \cite{Crab_MAGIC20}; (5) \cite{Crab_LHAASO21}; (6) \cite{VelaJr_HESS18}; (7) \cite{VelaJr_Fermi11}; (8) \cite{J1713.7_HESS18}; (9) \cite{J1713.7_Fermi11}.}
    \label{tab:Tier3}
\end{table*}

\begin{table*}[t]
    \centering
    \small
    \begin{tabular}{l p{6cm} l}
        \multicolumn{3}{c}{\textbf{Exclusions/Rejections}} \\
        \hline
        \textbf{Name} & \textbf{Exclusion Reason} & \textbf{References} \\
        \hline \hline
        G166.0+4.3 & possibly hadronic but sub-TeV & 1,2\\
        G73.9+0.9 & possibly hadronic but sub-TeV & 2,3\\
        HB9 (shell) & possibly hadronic but sub-TeV & 4,5,2\\
        MSH 15-56 & possibly hadronic but sub-TeV & 6,7,2\\
        HB 21 & possibly hadronic but sub-TeV & 8 \\
        G20.0-00.2 & possibly hadronic but sub-TeV & 9 \\
        Monoceros Loop & possibly hadronic but sub-TeV & 10,2 \\
        MSH 11-54 & possibly hadronic but sub-TeV & 11,12\\
        PKS 1209-51/52 & possibly hadronic but sub-TeV & 13\\
        G357.7-00.1 & possibly hadronic but sub-TeV & 14,15\\
        G150.3+4.5 & possibly hadronic but sub-TeV & 16 \\
        HB3 & possibly hadronic but sub-TeV & 17\\
        Kes 41 & possibly hadronic but sub-TeV & 18\\
        Kes 78 & possibly hadronic but sub-TeV & 19 \\
        G359.1-0.5 & possibly hadronic but sub-TeV & 20,21,22\\
        G106.3+2.7 & leptonic with low flux & 23,24,25 \\
        CTA 1 & leptonic with low flux & 26,27,28,29 \\
        3C 58 & leptonic with low flux & 30,31,32 \\
        CTB 37B & leptonic with low flux & 33,34,35 \\
        RCW 86 & leptonic with low flux& 36,37 \\
        G292.2-0.5 & leptonic with low flux& 38,39 \\
        J1731-347 & leptonic with low flux& 40,41 \\
        G54.1+0.3 & leptonic with low flux & 42,43,44\\
        G323.7-01.0 & uncertain SNR & 45\\
        G355.4+00.7 & classified SNR, insufficient spectral data & 46\\
        \hline \hline
    \end{tabular}
    \caption{SNRs (and uncertain SNRs) that were considered in this work and excluded from all tiers, as well as their respective reason for exclusion. Sources that are listed as \textit{possibly hadronic but sub-TeV} are those with apparently steep spectral indices, but that have visible exponential cutoffs or highly constraining upper limits at energies lower than $\sim 1$ TeV. Those that are listed as \textit{leptonic with low flux} are sources that were tested for inclusion in the Tier 3 catalog, but found to have an insufficient theoretical hadronic component when compared to the average integrated flux of sources in Tier 1. The single source listed as \textit{uncertain SNR} would be an interesting source to consider for this analysis, but is not a confirmed SNR. Lastly, the source listed as \textit{classified SNR, insufficient spectral data} is listed as a confirmed SNR in the First Fermi-LAT SNR catlog, meaning it is $\gamma-$ray bright, but a more detailed follow-up of the source in the GeV-TeV band has not been done to our knowledge. \\
    References: (1) \cite{G166.0_fermi13}; (2) \cite{suzuki+22}; (3) \cite{G73.9_fermi16}; (4) \cite{HB9_fermi22}; (5) \cite{HB9_suzaku+fermi19}; (6) \cite{MSH15-56_fermi18}; (7) \cite{MSH15-56_HESS18}; (8) \cite{HB21_fermi19}; (9) \cite{FermiSNRCat16}; (10) \cite{monoceros_fermi16};
    (11) \cite{MSH11-54_Tanaka13}; (12) \cite{MSH11-54_Fermi25};
    (13) \cite{PKS1209_Fermi24}; (14) \cite{MSH17-39_fermi14}; (15) \cite{MSH17-39_fermi13}; (16) \cite{G150_fermi24}; (17) \cite{HB3_fermi24}; (18) \cite{kes41_fermi18}; (19) \cite{kes78_fermi25}; (20) \cite{G359.1_fermi16}; (21) \cite{G359.1_HESS12}; (22) \cite{G359.1_fermi11};
    (23) \cite{G106.3_24}; (24) \cite{G106.3_ASgamma21}; (25) \cite{G106.3_Fermi19}; 
    (26) \cite{CTA1_LHAASO25}; (27) \cite{CTA1veritas13}; (28) \cite{CTA1_Fermi24}; (29) \cite{CTA1_Fermi16};
    (30) \cite{3C58_MAGIC14}; (31) \cite{3C58_FGL13}; (32) \cite{3C58_2FGL13};
    (33) \cite{zeng+17}; (34) \cite{CTB37B_Fermi16}; (35) \cite{CTB37B_HESS08};
    (36) \cite{RCW86_Fermi16}; (37) \cite{RCW86_HESS18};
    (38) \cite{acero+13PWNe}; (39) \cite{kargaltsev+13};
    (40) \cite{J1731_fermi+HESS17}; (41) \citep{J1713.7_fermi15};
    (42) \cite{G54.1_fermi25}; (43) \cite{G54.1_VERITAS+fermi18}; (44) \cite{cao+24_LHAASOCat};
    (45) \cite{G323.7_HESS17}; (46) \cite{FermiSNRCat16}.}
    \label{tab:exclusions}
\end{table*}

\section{Results}
\label{section:results}

The neutrino flux of any individual SNR is expected to be too faint to provide a detection with IceCube, but a stacked signal may be detectable in aggregate.  
We first calculate the expected individual-source sensitivity with IceCube using the 10 year public data of tracks only (\cite{IceCube_data21}; \url{https://doi.org/10.7910/DVN/VKL316}).
This is done using the \texttt{SkyLLH} software package developed by \cite{Wolf19}. 
The sensitivity is defined as the number of signal events (or flux) that it takes for the test statistic (TS) to be greater than the median of the background TS distribution 90\% of the time.
With \texttt{SkyLLH}, we use 10,000 background trials to calculate the TS threshold and compare them with signal trials--- injected ``signal-like" events obtained from the source's astrophysical parameters and the instrument response function (IRF) parameterizations, which account for the detector's intrinsic angular uncertainty and energy smearing. 
The expected signal for each source is dependent upon the source declination, spectral index, and in some cases the expected spectral cutoff.
All inputted values for SNRs in Tier 1 and Tier 2 are taken from Tables \ref{tab:Tier1} and \ref{tab:Tier2} respectively.

\begin{figure}
    \centering
    \includegraphics[trim = 7 7 7 5, clip=true, width=0.98\linewidth]{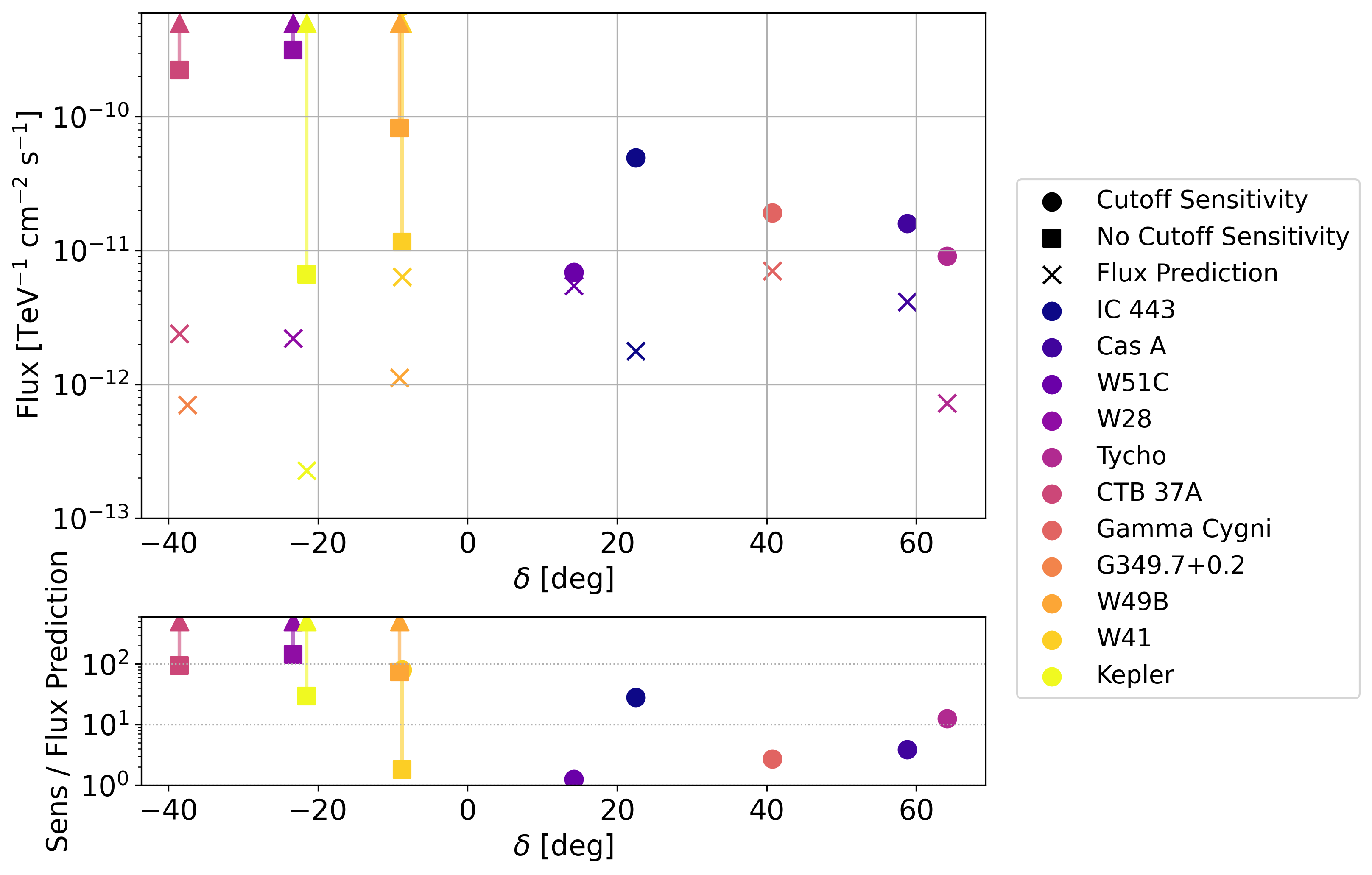}
    \caption{
    The predicted sensitivity of IceCube for each of the sources in the Tier 1 catalog as a function of source declination. Circles represent the flux at 1 TeV required for a given source to be detected by IceCube assuming the source has an exponential cutoff. Crosses represent the predicted true neutrino flux of the source at 1 TeV. The five sources without observed cutoffs (see Table \ref{tab:Tier1} (b) values) are also fit optimistically assuming a power law with no cutoff, represented here with squares, which provide a lower limit to the possible true sensitivity (see text). The bottom plot gives the ratio of the necessary flux to be observed by IceCube over the true flux of the source, where a ratio of $\sim 1$ indicates the source could potentially be detected individually, and ratios $\gg 1$ indicate the source is extremely unlikely to be detected individually. The source G349.7+0.2 has a sensitivity $\sim 9\times 10^{-5}$ TeV$^{-1}$cm$^{-2}$s$^{-1}$, which is outside the bounds of the figure, and a correspondingly high sensitivity to flux ratio which is also outside the figure's bounds.}
    \label{fig:Tier1_sensitivity}
\end{figure}

Figure \ref{fig:Tier1_sensitivity} shows the results for the expected IceCube sensitivity for each individual source in the Tier 1 catalog as a function of the source declination.
The five Tier 1 sources without observed cutoffs (namely: W28, CTB 37A, W49B, W41, and Kepler) are fit both with an exponential cutoff at their last data point (excluding upper limits), and with a power law and no cutoff. 
The remaining six sources in Tier 1 have observed cutoffs and are therefore fit only once, accounting for the exponential cutoff (see Table \ref{tab:Tier1} for details).
Circular markers represent the flux at $1$ TeV required for a given source to be individually detected by IceCube assuming the source has an exponential cutoff.
Square markers represent the same, but for a power law with no cutoff.
For the five sources fit both ways, the range between the upper and lower sensitivities bounds the most optimistic and pessimistic possible required neutrino fluxes to produce a detection with IceCube.
The cutoff sensitivities of these sources lie outside the bounds of the figure, and thus the square markers represent the lower limit sensitivities for each.
Crosses show the predicted true flux of the source, also at $1$ TeV.
The ratio of the sensitivity to the predicted flux is shown in the lower panel, where a ratio $\lesssim 1$ means the source may be individually detectable, and ratios $\gg 1$ mean the source is highly unlikely to be individually detected.
As in the top panel, the five sources without observed cutoffs have a range of possible sensitivity to predicted flux ratios, and we show here the lower limit.

We find that most sources in Tier 1 have sensitivity to flux ratios between $\sim 1$ and $100$, where the SNR W51C may even be individually detectable.
As expected, sources at very low declination are typically worse candidates for IceCube than those in the northern sky, and the SNR G349.7+0.2 is especially ill-suited with a calculated sensitivity of $\sim 9 \times 10^{-5}$ TeV$^{-1}$cm$^{-2}$s$^{-1}$ at 1 TeV, which is outside the bounds of the figure (the corresponding ratio in the lower panel also lies outside the figure's bounds) and is expected to make a negligible contribution to a stacked analysis.
Other sources have ratios $\sim 10$, which are ideal for stacking and demonstrate the potential power of this technique for this source class. 
A more careful estimate of the stacked sensitivity would add weights to sources depending on declination, but as an approximate method, we calculated the stacked sensitivity as,
\begin{equation}
    \Phi^{90\%}_{\rm stacked} \approx \frac{\Phi^{90\%}_{\rm med}}{\sqrt{N}} 
    \label{eq:stack}
\end{equation}
where $\Phi^{90\%}_{\rm med}$ is the median $90\%$ confidence value of the calculated sensitivities for Tier 1, $N$ is the number of sources, and thus $\Phi^{90\%}_{\rm stacked}$ is the $90\%$ confidence value of the stack of sensitivities. 
From our results, we obtain $\Phi^{90\%}_{\rm stacked} \approx 1.9\times 10^{-11}$ TeV$^{-1}$cm$^{-2}$s$^{-1} / \sqrt{11} \approx 5.8 \times 10^{-12}$ TeV$^{-1}$cm$^{-2}$s$^{-1} $. 
We then compare this value to the median predicted flux of our Tier 1 sample, which is found to be $\sim 2.2 \times 10^{-12}$ TeV$^{-1}$cm$^{-2}$s$^{-1}$, making the ratio of stacked sensitivity to median predicted flux $\sim 2.6$ in the most optimistic case.
In the most pessimistic case (and excluding G349.7+0.2), $\Phi^{90\%}_{\rm stacked} \approx 1.1 \times 10^{-10}$ TeV$^{-1}$cm$^{-2}$s$^{-1} $, leading to a ratio of $\sim 41$.
Ratios $\sim 1$, like our optimistic case, indicate that the stacked sources are near (or slightly below) the detection threshold for IceCube, and a ratio of $\sim41$ for our pessimistic case is less likely to be detectable.
However, we note that these estimates are only approximate and could be improved by more careful treatment with source weighting as a function of the SNR's declination. 
Adding additional years of track data and the inclusion of cascade events would also improve sensitivity.

\begin{figure}
    \centering
    \includegraphics[trim = 7 7 5 5, clip=true, width=0.98\linewidth]{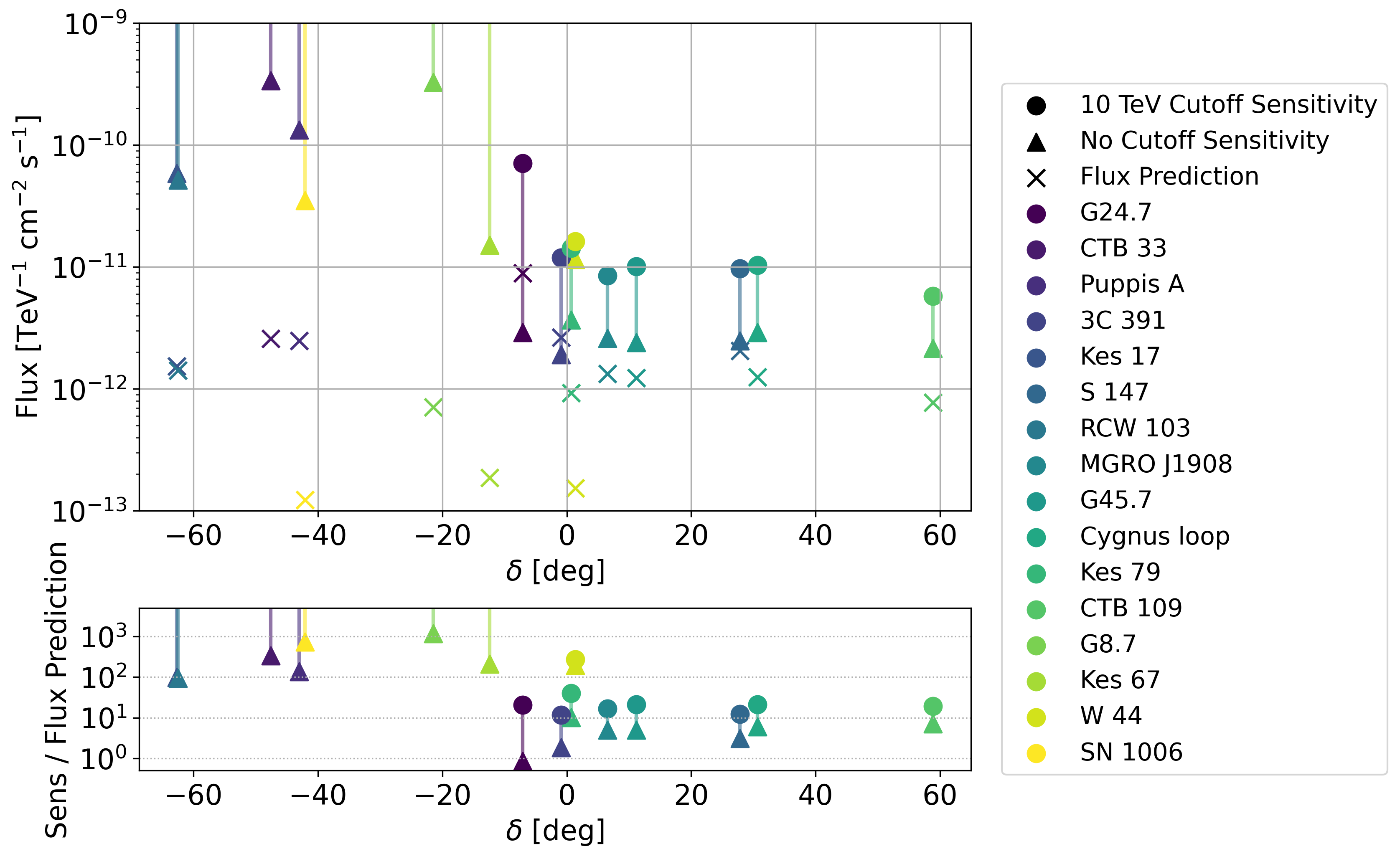}
    \caption{The same as Fig. \ref{fig:Tier1_sensitivity} but for Tier 2 sources. Sensitivities are calculated in two scenarios: one where all fits assume a power law with an exponential cutoff at 10 TeV (circles), and another where the fit is assumed to have no cutoff (triangles).}
    \label{fig:Tier2_sensitivity}
\end{figure}

The sensitivity results for Tier 2 can be found in Figure \ref{fig:Tier2_sensitivity} following the same procedure as discussed above for Tier 1. 
The most optimistic sensitivity estimate comes from treating Tier 2 sources as power laws with no cutoff (triangles), but we have also considered a more realistic scenario where all sources have an exponential cutoff at a representative $E_{\rm cut}$ approximately equal to the average value found among our Tier 1 sample, $10$ TeV (circles).
We find similar patterns to the sensitivity results in Tier 1, with greater sensitivity to sources in the northern sky compared to those in the south, and with some sources (such as G24.7+0.6) potentially individually detectable. 
In the scenario where all sources have no cutoff, the stacked sensitivity for all $16$ sources using Equation \ref{eq:stack} is $\Phi^{90\%}_{\rm stacked} \approx 1.9 \times 10^{-12}$ TeV$^{-1}$cm$^{-2}$s$^{-1} $ and the median predicted flux is $\approx 1.3\times 10^{-12}$ TeV$^{-1}$cm$^{-2}$s$^{-1} $. 
This gives a ratio of stacked sensitivity to median predicted flux of $\sim 1.5$.
In the less optimistic scenario where all sources are assumed to have a cutoff at $10$ TeV, the ratio of stacked sensitivty to median predicted flux is $\sim8.4$. 
Similar to the discussion for the Tier 1 catalog, this ratio could potentially be improved by a more careful treatment of source weighting, additional years of track data, and the inclusion of cascade data. 
However, we also note that our Tier 2 sources are treated optimistically, and future $\gamma-$ray observations may reveal that some sources have cutoffs even below a TeV.
Whether these competing effects tend more towards improving sensitivity or worsening it is difficult to say a priori, but a non-detection of this catalog from an internal IceCube analysis could be used as evidence that SNRs in this catalog tend to have steep energy drop-offs around a TeV, supplementing $\gamma$-ray observations.

The Tier 3 catalog contains only three sources (the Crab, Vela Jr, and J1713.7) all of which are expected to be dominated by leptonic emission, but which could contain an under-dominant hadronic component with appreciable neutrino flux compared to sources in Tier 1. 
Both Vela Jr and J1731.7 lie in the southern hemisphere where IceCube is less sensitive, and thus have individual sensitivity to predicted flux ratios for the assumed hadronic component that are $>10^3$, and we therefore conclude that these sources are not expected to contribute meaningfully to a stacked signal. 
The Crab, however, was found to have an individual ratio $\sim 2.8$ for its assumed maximal hadronic component, which could make it an interesting, albeit speculative, target.
More concrete evidence of proton-proton collisions in the vicinity of the Crab may help justify the inclusion of this source in a stacked analysis.

Lastly, we provide an estimate of the stacked sensitivity for Tiers 1 and 2 combined using the most optimistic sensitivities from Tier 1, first in the case where Tier 2 is assumed to have a cutoff at 10 TeV: 
$\Phi^{90\%}_{\rm stacked} \approx 3.7 \times 10^{-12}$ TeV$^{-1}$cm$^{-2}$s$^{-1} $ and the median predicted flux over both tiers is $\approx 2.3\times 10^{-12}$ TeV$^{-1}$cm$^{-2}$s$^{-1} $, leading to a ratio of sensitivity to predicted flux of $\sim 1.6$.
In the most optimistic scenario where all Tier 2 spectra are assumed to have no cutoff, $\Phi^{90\%}_{\rm stacked} \approx 2.9 \times 10^{-12}$ TeV$^{-1}$cm$^{-2}$s$^{-1} $, leading to a ratio of sensitivity to predicted flux of $\sim 1.3$.
While our procedure for estimating the stacked sensitivity is approximate, and thus caution that the precise values reported would likely change in a more careful log likelihood analysis, we  
nonetheless argue that this test motivates a followup of the combined stack of Tier 1 and 2 sources by IceCube.

\subsection{Contamination of Leptonic Sources}
\begin{figure}
    \centering
    \includegraphics[width=0.98\linewidth]{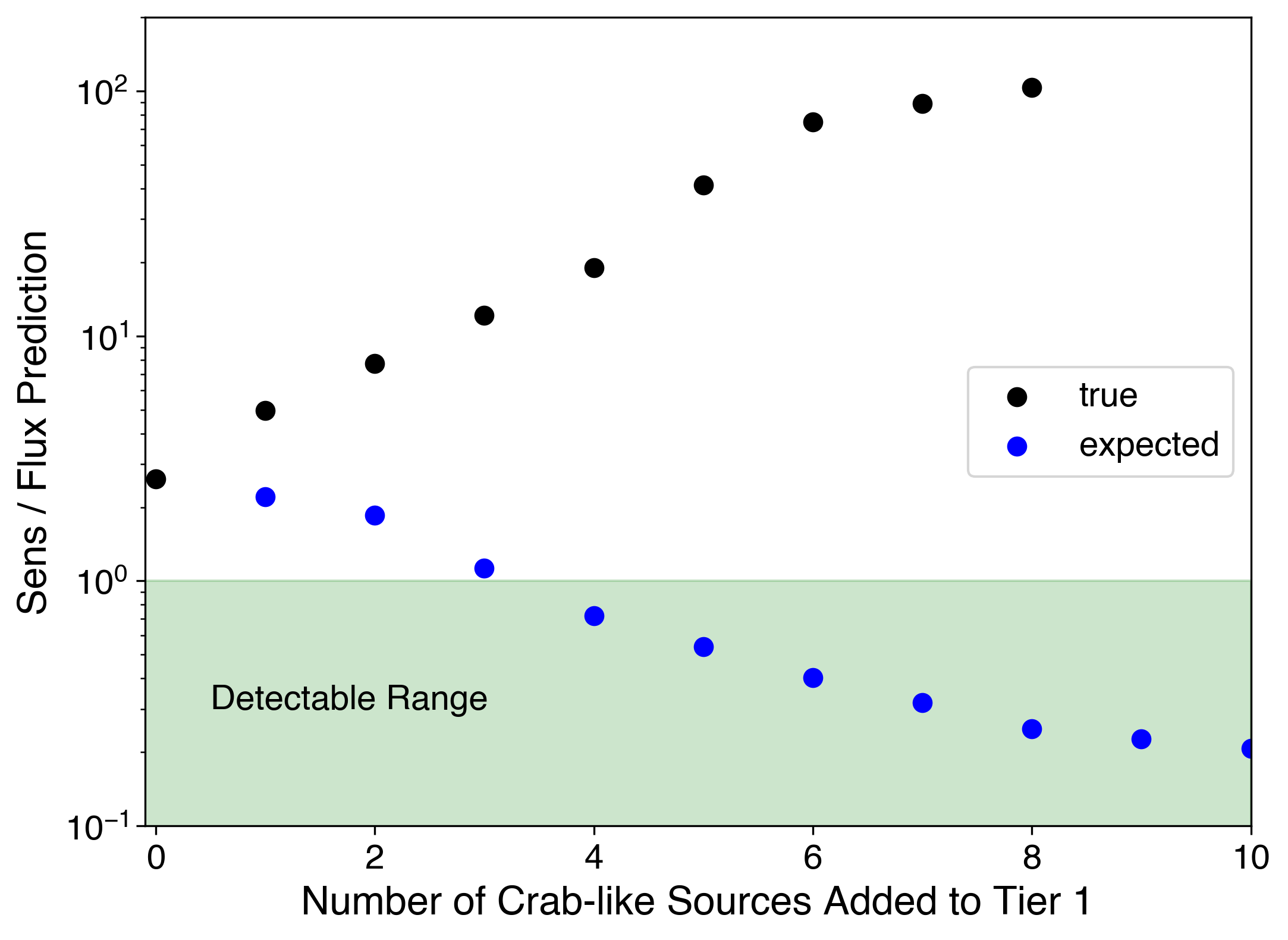}
    \caption{The stacked sensitivity-to-predicted-flux ratio of the Tier 1 catalog with increasing contamination from leptonic ``Crab-like" sources. Black circles represent the true stacked ratio, and blue circles represent the expected ratio one would get if they erroneously assume a 1:1 ratio of $\gamma-$rays to neutrinos for a leptonic source. The green, shaded region represents ratios in the detectable range for IceCube.}
    \label{fig:leptonic_contamination}
\end{figure}

The previous section demonstrates the benefit of stacking hadronic sources, but we now comment on the \textit{penalty} of including leptonic sources into the stack.
While many leptonic sources are extremely bright at TeV and even PeV energies \citep{cao+24_LHAASOCat}, this does not necessarily imply that their respective neutrino fluxes are also large;
inverse Compton radiation can produce $\gamma-$rays at these energies without any corresponding neutrinos.
The inclusion of leptonic sources into a stacked sample would increase the expected signal, which could not be detected.

Figure \ref{fig:leptonic_contamination} shows the effect on the ratio of the stacked sensitivity to the median flux prediction for our Tier 1 sample of hadronic SNRs, but with increasing contamination by ``Crab-like" sources--- sources with very high $\gamma-$ray fluxes and hard spectral indices, but which we assume are fully leptonic and contribute no neutrinos. 
For each added ``Crab", we calculate the new stacked sensitivity that would be naively expected if the $\gamma-$ray to neutrino ratio is 1:1, and use this to calculate the expected ratio of sensitivity to predicted flux of the stack (blue circles).
At the same time, we calculate the true stacked sensitivity assuming the added source contributes no neutrinos, and use this to calculate the true sensitivity to predicted flux ratio (black circles).
The trend in the expected flux shows the naive assumption that prospects for detection will \textit{improve} with the inclusion of more of these sources, while the true ratio shows it is actually worsening the prospects considerably. 
The stacking procedure used in this work allows us only to approximate this effect, but these results are a qualitative argument that the inclusion of leptonic sources can be detrimental, particularly if the desired signal is on the cusp of detectability to begin with.

\section{Conclusions}
\label{section:conclusion}

We have considered all known Galactic SNRs and categorized them into three tiers based on their likelihood to produce appreciable neutrino emission within the observable energy range for IceCube (see Figure \ref{fig:flowchart}).

Tier 1 consists of 11 SNRs with steep  GeV-TeV spectra (indicating that their $\gamma$-ray emission is likely hadronic), and that are observed to have emission at energies $> 1$ TeV where IceCube becomes sensitive to astrophysical neutrinos.
These sources are then fit with a power law (either with or without an exponential cutoff), in order to predict their expected neutrino signal.

Tier 2 contains 16 likely hadronic sources with steep GeV spectra, but which are worse candidates for IceCube due to a lack of TeV observations.
For these sources, we have extrapolated the flux at high energies in the optimistic case of a continuous power law spectrum either with no cutoff, or with a cutoff at $10$ TeV, in order to make predictions about the possible neutrino signal.

Finally, Tier 3 is comprised of sources which are spectrally hard and thus have emission which is likely dominated by leptonic processes (namely, Inverse Compton), but are especially bright such that a substantial hadronic component could be hidden beneath the total spectrum. 
We found only three sources which could contribute an appreciable neutrino flux from an under-dominant hadronic component, and estimated the neutrino flux assuming an $E^{-2}$ spectrum. 

Our results show that Tier 1 has $\mathcal{O}(1)$ sources which may be detectable individually (in particular, W51C), and that the stack of Tier 1 sources has an optimistically calculated sensitivity to flux prediction ratio of $\sim 2.6$ using only IceCube 10-year track data which has been publicly released and without weighting sources by declination. 
Our optimistic results for Tier 2 similarly produce a sensitivity to flux prediction ratio $\sim 1.5$ using our basic estimation procedure.
The rough stacked sensitivity estimate for the combined Tier 1 and Tier 2 catalogs gives a ratio $\sim 1.3$ using the most optimistic fit values, indicating that the combined catalogs may be a particularly promising route towards detecting a neutrino signal. 
Our results emphasize the potential of a followup using additional years of track data and including cascade events, which could be important for going beyond a detection and reaching $3\sigma$ evidence potential.

Opening the neutrino window onto SNRs has the potential to further our understanding of the sources of Galactic CRs and, in particular, to test whether SNRs are capable of accelerating particles to the CR ``knee" around a PeV and beyond. 
This work presents followup targets that are motivated by theory, specifically rejecting sources likely to be leptonic, in order to have the best possible chance of detecting a neutrino signal in a stacked analysis.
It is important to mention that $\gamma-$ray astronomy is an expanding field, and that future instruments like the Cherenkov Telescope Array (CTA) \citep{CTA_whitepaper19}, and additional LHAASO data may alter the classification of any individual SNR, causing sources to shift among the catalogs presented in this work, downgrading some and upgrading others.
Despite this, the \textit{strategy} presented in this work of categorizing sources based on their likelihood of hadronic emission (using the most up-to-date data available and basic theory of DSA and radiation processes)  represents the best possible chance of detecting neutrino emission from any group of SNRs.

Lastly, we note that the present catalogs were constructed with IceCube in mind, but that there is strong motivation to extend this framework to other neutrino observatories sensitive in the TeV–PeV range but covering complementary regions of the sky.
In particular, a corresponding analysis using KM3NeT’s ARCA telescope \citep{KM3NeT_sciencebook19} would be highly valuable, given its potential superior sensitivity to the Southern hemisphere, where roughly half of our source sample resides.
Future multimessenger studies combining $\gamma$-ray, neutrino, and CR observations will be essential to fully characterize particle acceleration in SNRs and to resolve the long-standing question of the origin of the highest-energy Galactic cosmic rays.

\section{Acknowledgments}
The authors thank Chiara Bellenghi for her help updating the \texttt{SkyLLH} software to include exponential cutoffs.
They also thank Shuo Zhang, Steve DiKerby, and Kayla Owens for helpful discussions.
E.S.~was partially supported by the NSF Graduate Research Fellowship Program (grant 2140001) and the IL Space Grant Consortium Fellowship (ILSGC).
D.C.~was partially supported by NASA (grants 80NSSC24K0173 and 80NSSC23K1481) and NSF (grants AST-2510951 and AST-2308021).


\bibliography{main}{}
\bibliographystyle{aasjournal}



\end{document}